\documentclass[twocolumn,a4paper,14pt]{revtex4}
\usepackage[english]{babel}
\usepackage{epsfig}
\usepackage{latexsym}
\usepackage{float}
\usepackage{epsfig}
\usepackage{here}
\def\be{\begin{equation}}
\def\ee{\end{equation}}
\def\bea{\begin{eqnarray}}
\def\eea{\end{eqnarray}}

\begin{document}
\title{A new technique for the sp$^2$/sp$^3$ characterisation of carbon materials}
\author{J.~T.~Titantah and D. Lamoen\\
{{\small \it TSM}, \small \it  Department of Physics, University of 
Antwerp, Groenenborgerlaan 171, 2020 Antwerpen, Belgium}\\
\date{\normalsize (\today)}}
\begin{abstract}

We present a technique to determine the sp$^3$/sp$^2$ ratio of carbon 
materials which is based on the electron energy-loss spectroscopy and which 
uses the theoretical spectrum of graphite obtained from ab initio electronic 
structure calculations. The method relies on the separation of the  $\pi^*$ 
and $\sigma^*$ components of the carbon K-edge of graphite. This $\pi^*$ 
spectrum is adopted and assumed to be transferable to other carbon systems 
given an appropriate parametrisation of the broadening. The method is applied on a series of Monte-Carlo generated amorphous carbon structures and is shown to be  stable over a wide range of the energy windows for which spectral integration is performed. The results are found to be in good agreement with the sp$^3$ fraction obtained from a microscopic scheme which uses the $\pi-$orbital axis vector (POAV1) analysis. The technique was also applied on a series of experimental spectra of amorphous carbon and was found to be in good agreement with the results obtained from a functional fitting approach. 

\end{abstract}
\baselineskip=6mm
\maketitle
One of the parameters that determine the hardness of amorphous carbon is the 
fraction of sp$^3$ bonded carbon atoms. This fraction can be determined by 
making use of the X-ray photoelectron spectroscopy (XPS)~\cite{haerle01}, the 
plasmon energy technique~\cite{egerton86,ferrari00}, the  near-edge X-ray 
absorption spectroscopy (NEXAFS)~\cite{comelli88} and the high-energy 
electron energy-loss spectroscopy (HEELS)~\cite{ferrari00,fallon93, bruley95,papworth00,berger88}.  

The high-energy EEL spectrum of an amorphous carbon is made of a $\pi^*$ 
part whose onset is at about 284 eV and a $\sigma^*$ component whose onset is 
 about 5-7 eV from that of the $\pi^*$ component. The HEELS sp$^3$/sp$^2$ 
characterisation technique is based on the ability to isolate the $\pi^*$ 
features and subsequently its cross-section from the entire spectrum. All 
the isolation methods that are often used are limited by the assumption 
that the $\pi^*$ and the $\sigma^*$ bands are well separated in energy 
such that their cross-sections can be independently deduced by spectral 
integrations over fixed energy windows. But it is well known that plural 
inelastic scattering such as core-loss excitation plus additional plasmon 
losses modifies the edge intensities by spreading oscillator strengths 
to higher energies~\cite{bruley95}. The $\pi^*$ band can spread  
to energies as high as 30 eV beyond the edge onset thereby 
overlapping with the $\sigma^*$ band. 
Density functional theory (DFT) calculations on graphite and amorphous carbon, 
which will be presented elsewhere~\cite{titantah-sp3}, reveal that 
the $\pi^*$ band is exceptionally broad, extending even to energies beyond 
30 eV. Therefore, a sp$^3$/sp$^2$ ratio characterisation that ignores these
 finite $\pi^*$ intensities (of the HEELS or NEXAFS) at higher energies is 
not entirely consistent. In this letter we present a new technique for the 
characterisation of the sp$^3$/sp$^2$ ratio in (amorphous) carbon systems 
which is based on the $\pi^*$/$\sigma^*$ separation of the calculated 
energy-loss near-edge structure (ELNES) of graphite. We show that this 
consistent treatment of the $\pi^*$ and the  $\sigma^*$ spectra can lead 
to sp$^3$ proportions that are independent of the choice of the size of the 
integration energy window in sharp contrast with other 
techniques~\cite{bruley95,papworth00}.

DFT calculations are performed using the all-electron 
Full-Potential-Linearised-Augmented-Plane-Wave (FLAPW) code 
WIEN2k~\cite{wien2k}. The exchange and 
correlation energy is treated using the local density 
approximation~\cite{perdew92}. Muffin-tin radii (R$_{MT}$) are fixed at 1.3 
atomic units (a.u.) for graphite and 1.25 a.u. for amorphous carbon, while 
R$_{MT}\times$K$_{max}$ values of 5.5 and 5.0 are 
used for graphite and amorphous carbon, respectively; K$_{max}$ being the  
plane wave cut-off.  Up to 1000 k-points are used to sample the 
full Brillouin zone of graphite (72 in the irreducible Brillouin zone (IBZ)) 
while 20 (4 in the IBZ) are used for the amorphous carbon. The core-hole effect is 
introduced in the ELNES of graphite via the so-called excited-core method (in 
a supercell approach). The $\pi^*$ component of the ELNES of graphite is 
obtained using the expressions of ref.~\cite{nelhiebel99} for the 
differential cross-section which was 
recently implemented on hexagonal boron nitride~\cite{hebert00} and on 
nanotubes~\cite{titantah-nano}.

In order to obtain insight into the local structure of amorphous carbon, we 
generate amorphous carbon systems by means of classical Monte Carlo (MC) 
simulations based on the Tersoff empirical potential~\cite{tersoff88a} for 
carbon \cite{tersoff88b}. We generated amorphous carbon structures at densities 
of 2.0, 2.6, 3.0 and 3.6 g/cm$^3$  by quenching hot gaseous carbon from 8000K 
down to 300K. This was done via a continuous space, periodic boundary 
condition (PBC), constant volume, constant temperature and constant number of 
atoms classical Metropolis Monte Carlo procedure on cubic cells made of 64 
atoms. The ELNES spectra (without core-hole) were computed for the four 
generated carbon systems. The average (over all atoms) ELNES spectra of each 
of the systems were calculated. On 
Figure~1 we show the ELNES spectra of the four 
generated carbon systems. 

\begin{figure}[H] 
\begin{center}
\includegraphics[width=2.5in,height=2.5in,angle=-90]{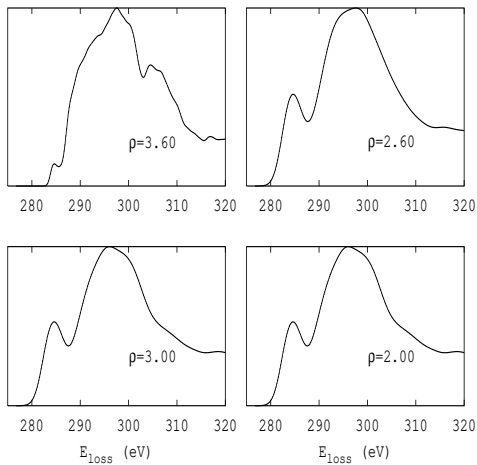}
  \caption{Total ELNES (in the sudden approximation) of four MC-generated carbon structures of varying densities of 2.00-3.60 g/cm$^3$.}
\end{center}
\label{fig:tot-elnes}
\end{figure}
The simulated 
structures allow us to look at the local environment of every atom and to 
determine the sp$^3$ fraction from a microscopic point of view. For curved 
carbon systems like the fullerenes and the nanotubes, the effect of curvature 
on the hybridisation state of carbon atoms has been successfully described by 
using the $\pi-$orbital axis vector (POAV1) analysis~\cite{haddon88,haddon86a,haddon86b,haddon99}. The application of this technique to all three-coordinated atoms of
 our MC-generated  structures provides a more precise approach to characterise 
the sp$^3$ fraction because it accounts for the non-planarity of the 
three-coordinated atoms. However, not all four-coordinated atoms are having the ideal sp$^3$ local bonding structure as follows from the local geometry and our calculations of the core-loss ELNES spectra of the generated structures.  We can easily account for the distortion-induced sp$^2$ character in such four-coordinated atoms by analysing the $\pi^*$ components for every non ideal tetrahedral four-coordinated atom  as explained in ref.~\cite{titantah-sp3}. The sp$^3$ fraction obtained by this approach is consistently higher than that obtained by using the coordination number method for low and intermediate densities (see Table~\ref{tabl:sp3}). For high densities the two methods yield the same results.

The technique of the sp$^3$/sp$^2$ characterisation presented here is based on 
the assumption that the ELNES of any carbon atom can be decomposed into two 
independent parts. One of these ELNES components 
(the $\pi^*$ component) is sought by performing first-principles calculations 
on graphite for which the two components are well defined. To some extent, the method is comparable to the experimental work of ref.~\cite{browning91}. 
The sp$^3$/sp$^2$ quantification procedure is described as follows
\begin{itemize}
\item We identify the energy position E$_\pi$ of the main $\pi^*$ peak 
from the spectrum of the 
uncharacterised sample and rescale the intensity to have a value of one 
at this peak. We denote this new spectrum as I(E) where E is the energy loss.
\item  We consider a normalised smoothening function $g(\sigma,E)$ ( which 
could be a Gaussian or a Lorentzian) of full width at half maximum $\sigma$.
\item We define an energy scale shift parameter $s$ by which the unbroadened 
calculated graphite $\pi^*$ spectrum should be shifted to align the maximum 
of its main 
$\pi^*$ peak to that of the unknown sample. After the spectral shift the 
$\pi^*$ spectrum is scaled to one at E$_\pi$. 
\item Using this new shifted graphite $\pi^*$ spectrum which is convolved with $g(s,\sigma,E)$, denoted as $\Pi(s,\sigma,E)$, we define the function
\begin {equation}
F(s,\sigma)=\int_{E_{min}}^{E_\pi} \left(I(E)-\Pi(s,\sigma,E)\right)^2 dE
\label{eq:min}
\end{equation}
where E$_{min}$ is the edge-onset. In Eq.(\ref{eq:min}) it is implied that the spectrum of the amorphous carbon is entirely of $\pi^*$ character up to the main $\pi^*$ 
peak. 
\item F is minimised for $s$ and $\sigma$. The minimal values of $s$ and 
$\sigma$ are used for $\Pi(s,\sigma,E)$ to deduce the $\pi^*$ spectrum 
which best fits the lower energy part of the unknown spectrum. This shifted and convolved $\pi^*$ spectrum is assumed to be that of the unknown amorphous carbon system. 
\item An energy window $\Delta E$ whose lower limit is fixed at the edge onset 
is chosen and the integrals $I_\pi(\Delta E)=\int_{E_{min}}^{E_{min}+\Delta E} \Pi(s,\sigma,E) dE$ and $I_{Total}(\Delta E)=\int_{E_{min}}^{E_{min}+\Delta E} I(E) dE$ are evaluated.
\item The ratios, for various $\Delta E$, $R(\Delta E)=I_\pi(\Delta E)/I_{Total}(\Delta E)$ are calculated.
\item The ratio $R_0(\Delta E)$ for a well characterised sample, whose sp$^2$ fraction is $x_0$ is known, is  also calculated  and the sp$^2$ fraction in the unknown amorphous carbon sample is deduced as 
\begin{equation}
x(\Delta E)={R(\Delta E)\over R_0(\Delta E)}x_0
\end{equation} 
and the sp$^3$ fraction is $1-x$.
\end{itemize}

To test and validate this method, we applied this technique to the ELNES spectra of the four generated amorphous carbon systems described above. The ratio $R(\Delta E)$ for each system is plotted as a function of $\Delta E$ on Figure~2. From this figure it is seen that for each energy window $R$ decreases monotonically as the density increases. It is also seen that for the wide range of $\Delta E$ considered ($15~eV<\Delta E<55~eV$) all the curves showing the variation of 
 $R$ with $\Delta E$ seem to be parallel to each other. This suggests that if one of the spectra was a reference spectrum then the ratio of sp$^2$ carbon in each spectrum with respect to this reference will be fairly constant over this wide energy range. This demonstrates that the method should yield sp$^2$ or sp$^3$ fractions that are independent of the choice of the energy window in contrast to other techniques. In those techniques the sp$^3$ fraction depends crucially on both the width and position of the energy window.
\begin{figure}[H]
\begin{center}
\includegraphics[width=2.5in,height=2.5in,angle=-90]{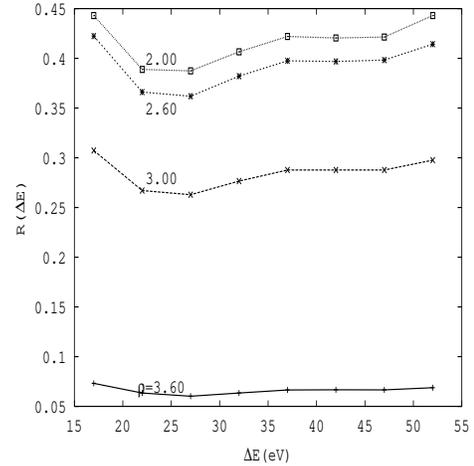}
  \caption{The ratio R=Area($\pi^*$)/Area($\pi^*$+$\sigma^*$) for the seven generated carbon structures. Remark that all the lines are approximately parallel to each other.}
\end{center}
  \label{fig:stability}
\end{figure}

To determine the sp$^3$ fraction, we take one of the Monte-Carlo generated amorphous carbon structures as a reference. Because the sp$^3$ fraction determined by the POAV1 analysis coincides with that obtained by the  nearest neighbour counting method for high densities (see Table~\ref{tabl:sp3}), the structure having a density of 3.6 g/cm$^3$ with an sp$^3$ fraction of 90\% is considered as the reference.  The resulting fractions for the other three systems are shown on Figure~3. The sp$^3$ fraction is clearly seen to be independent of the size of the energy window. The largest variation recorded did not exceed 6\%. The values obtained, reproduced in Table~\ref{tabl:sp3}, were in very good agreement with those obtained from a POAV1-based analysis. These results are given as intervals because of the standard deviations on the POAV1-corrected sp$^3$ fractions recorded on the entire systems of the three-coordinated atoms. In the table we also report the proportion of 4-coordinated atoms in each of the systems.
\begin{figure}[H]
\begin{center}
\includegraphics[width=2.5in,height=2.55in,angle=-90]{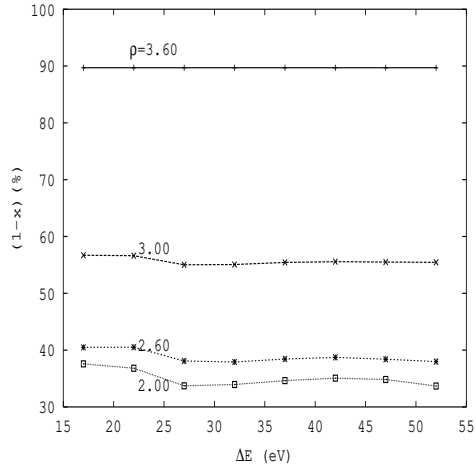}
  \caption{The sp$^3$ percentage for the four generated carbon structures. The $\rho=3.6 g/cm^3$ sample made of 10.3\% sp$^2$ atoms (according to the POAV1 approximation) is used as the reference system.}
\end{center}
  \label{fig:sp3}
\end{figure}

\begin{table}[H]
\begin{center}
\begin{tabular}{|c|c|c|c|}
\hline
density& new method&coordination number& POAV1\\
\hline
2.00& 32-38& 16& 28-32\\
\hline
2.60& 39-40& 20& 35-37\\
\hline
3.00& 56-58& 55& 54-56\\
\hline
3.60& 90& 90& 90\\
\hline
\end{tabular}
\end{center}
\caption{The sp$^3$ percentages of the four generated amorphous carbon systems.
The 3.6 g/cm$^3$ system is considered as reference and its sp$^2$ ratio (0.103) according to the POAV1 is adopted. }
\label{tabl:sp3}
\end{table}

The use of this technique is not restricted to computer generated samples (in contrast to the POAV1 analysis) but can also be applied to experimental spectra.  On Figure~4 and in Table~\ref{tabl:sp3-exp} we show the sp$^3$ fractions of five spectra obtained from ref.~\cite{Ann}. In this table we also listed the values obtained in~\cite{Ann} using a functional fitting method with three Gaussians. Remark the good agreement with the functional fitting technique. Sample I showed the strongest dependence on $\Delta E$, but for the large energy window range of 15~eV - 45~eV the variation did not exceed 10\% for this sample. 

\begin{figure}[H]
\begin{center}
\includegraphics[width=2.5in,height=2.55in,angle=-90]{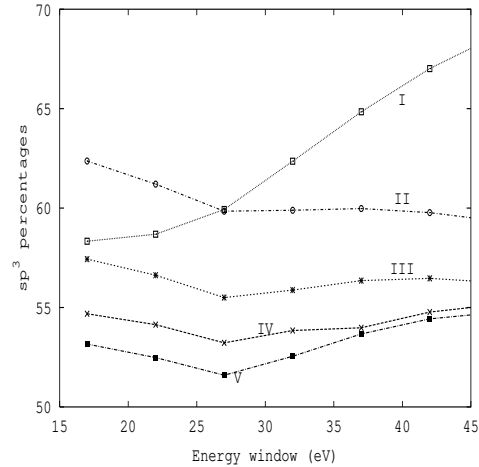}
  \caption{The sp$^3$ percentage for the five amorphous carbon samples.}
\end{center}
  \label{fig:sp3-exp}
\end{figure}
Table~\ref{tabl:sp3-exp} compares the sp$^3$ percentages of the five amorphous carbon samples with results obtained using the functional fitting method.
\begin{table}[H]
\begin{center}
\begin{tabular}{|c|c|c|c|}
\hline
Sample& new method&Fitting technique\\
\hline
I& 58-68&67-68\\
\hline
II& 60-63& 61\\
\hline
III& 56-57& 56\\
\hline
IV& 54-55& 51-53\\
\hline
V&52-55& 53\\
\hline
\end{tabular}
\end{center}
\caption{The sp$^3$ percentages of the five amorphous carbon samples.
The 3.6 g/cm$^3$ system is considered as reference and its sp$^2$ ratio (0.103) according to the POAV1 is adopted. Apart from sample I, all samples show excellently stable results within the 30 eV window.}
\label{tabl:sp3-exp}
\end{table}

In summary, We have developed a  method to quantify the sp$^3$ fraction of 
carbon materials based on the theoretical separation of the $\pi^*$ 
and $\sigma^*$ components of the carbon 1s electron energy-loss spectra.
The $\pi^*$ spectrum of graphite is adopted and assumed to 
be transferable to other carbon systems given  an appropriate parametrisation 
of the broadening. This has enabled the isolation of the $\pi^*$ spectrum of 
the carbon systems and a subsequent characterisation of their sp$^3$ contents.
We have applied the method to a series of Monte-Carlo generated 
amorphous carbon structures and 
shown that the method gives stable results over a wide range of the energy 
window used for spectral integration and that the sp$^3$ 
fractions obtained using this method agree with values resulting from an exhaustive analysis of the distorsion-induced hybridisation of every carbon atom by means of the $\pi-$orbital axis vector (POAV1) technique. The method has also been applied to 
five experimental amorphous carbon spectra and found to be in good agreement with the results obtained from a more traditional functional fitting technique.

We would like to thank the authors of ref.~\cite{Ann} for providing 
their data and K. Jorissen, A.-L. Hamon, J. Verbeeck, D. Schryvers and 
P. Potapov for stimulating discussions. Part of this work was supported by 
the Special Research fund of the University of Antwerp (BOF-NOI) and a GOA from the University of Antwerp ``Characterisation of nano-structures by means of advanced electron energy spectroscopy and filtering''.

\end{document}